\documentclass{eccm-ecfd}

\usepackage{graphicx}
\usepackage{amsmath}
\usepackage{amsfonts}
\usepackage{amssymb}
\usepackage{bm}
\usepackage[colorlinks=true,breaklinks=true]{hyperref}
\usepackage[numbers]{natbib}
\usepackage{doi}
\usepackage{url}

\usepackage{multicol}
\usepackage{gensymb}
\usepackage{mathtools}
\usepackage{epstopdf}
\usepackage{wrapfig}
\usepackage{caption}
\usepackage{subcaption}
\usepackage{mhchem}
\usepackage{bm}
\usepackage{multirow}

\newcommand*{\V}[1]{\expandafter\mathbf #1}
\renewcommand*{\v}[1]{\bm{\mathrm{#1}}}
\renewcommand{\vec}[1]{\bm{\mathrm{#1}}}

\allowdisplaybreaks

\title{MODELLING AND SIMULATIONS OF ELECTRICAL PROPAGATION IN TRANSMURAL SLABS OF SCARRED LEFT VENTRICULAR TISSUE}

\author{PETER MORTENSEN$^{1}$, MUHAMAD HIFZHUDIN BIN NOOR AZIZ, HAO GAO \& RADOSTIN D. SIMITEV$^{2}$}

\heading{Mortensen, Noor Aziz, Gao and Simitev}

\address{
School of Mathematics and Statistics, University of Glasgow Glasgow G12 8SQ, UK
\and
$^{1}$ P.Mortensen.1@research.gla.ac.uk, 
\and
$^{2}$ Radostin.Simitev@glasgow.ac.uk, ORCID ID orcid.org/0000-0002-2207-5789
}


\abstract{We report three-dimensional and time-dependent numerical
simulations of the propagation of electrical action potentials in a
model of rabbit ventricular tissue. The simulations are performed
using a finite-element method for the solution of the monodomain
equations of cardiac electrical excitation. The parameters of a
detailed ionic ventricular cell model are re-fitted to available
experimental data and the model is then used for the description of the
transmembrane current and calcium dynamics. A region with reduced
conductivity is introduced to model a myocardial infarction
scar. Electrical activation times and density maps of the 
transmembrane voltage are computed and compared with
experimental measurements in rabbit preparations with myocardial
infarction obtained by a panoramic optical mapping method.}

\begin{document}

\thispagestyle{empty}

\section{INTRODUCTION}
The heartbeat is controlled by a particular pattern of an electrical
wave. When the heart is damaged by a myocardial infarction (MI), this
pattern is disturbed, leading to arrhythmias
and heart failure. Thus, it is important to understand how this
pattern is formed and how MI scars affect it. Here we begin to explore
these effects by mathematical modelling and simulation of action potential
propagation in a slab of cardiac tissue, based on and compared to
experiments performed on post MI rabbit hearts.

\section{MATHEMATICAL MODEL}

\subsection{Tissue model}
We consider the monodomain model given by the set of equations
\begin{subequations}
\label{eq::monodomain}
\begin{gather}
\label{eq::tissue_monodomain} 
\chi C_m\frac{\partial V}{\partial t} - \nabla \cdot (\vec{\sigma} \nabla
V) = -\chi \, I_\text{ion}  -\chi \, I_\text{stim} (\vec{x},t),  \\ 
\label{eq::Iion}
I_\text{ion} = I_\text{ion}\big(V(\vec{x},t),\vec{y}(\vec{x},t)\big),\\
\label{eq::gates}
\frac{\partial \vec{y}}{\partial t} = \vec{R}\big(V,\vec{y}\big),\\
\text{for } \quad \vec{x} \in \Omega, \quad t \in [0,\infty),
\end{gather}
with boundary conditions 
\begin{gather}
\label{eq::bc_monodomain}
\frac{\partial V}{\partial \vec{n}} = 0 \quad \text{on } \quad \vec{x} \in \partial\Omega,
\end{gather}
\end{subequations}
in a spacial domain $\Omega\in \mathbb{R}^3$ representing a piece of
cardiac tissue with $\vec{n}$ being the outer normal unit vector to
its boundary $\partial \Omega$. 
Here $V$ is the cardiac transmembrane electric voltage potential measured in $\text{mV}$,  
$I_\text{ion}$ is electric current density across the membrane of cardiomyocyte cells measured in $\mu\text{A mm}^{-2}$, 
$I_\text{stim}$ is the density of an externally applied stimulus current also measured in
$\mu\text{A mm}^{-2}$, 
$\chi$ is the surface-to-volume ratio of cardiomyocytes measured in $\text{mm}^{-1}$, 
$\vec{\sigma}$ is the effective conductivity of the cardiac tissue measured in $\text{mS
  mm}^{-1}$ 
and $C_m$ is the specific cell membrane capacitance measured in $\mu\text{F mm}^{-2}$. 
The transmembrane current $I_\text{ion}$ is modelled as a function of
a vector of state variables $\vec{y}$ representing ionic
concentrations and ionic channel gating variables determined by a
system of nonlinear ordinary differential equations with rates given
by $\vec{R}$. 
The monodomain model provides a biophysical continuum representation
of cardiac electrophysiology in both space and time, linking
tissue-scale electrical propagation with cellular electrical
excitation.  
The monodomain equations are derived from the laws of conservation of
charge and the assumption that infinitesimal pieces of the
cardiomyocyte membrane may be modelled as an circuit of a conductor
and capacitor connected in parallel.

Specific values for $\sigma$, $C_m$ and $\chi$ as well as for the
geometry of the tissue used are given further below.

\subsection{Single-cell electrophysiology models}
A large number of single-cell ionic current models  
given by equations \eqref{eq::Iion} and \eqref{eq::gates} of the
monodomain system \eqref{eq::monodomain} exist to represent the
conducting properties of cardiac myocyte membranes.  
These models can be classified into conceptual
and detailed with the detailed ionic models further divided into
models for various type cells (atrial, ventricular, sino-atrial,
Purkinje), various species (human, porcine, canine, leporine, murine)
and various state of remodelling (healthy normal, in heart 
failure etc.) These models are subject to continuous re-evaluation and 
refinement as new experimental data becomes available. The
contemporary models include tens of ordinary differential equations and 
online model repositories such as CellML\footnote{\url{http://models.cellml.org}}
have been setup for ease of their dissemination and use.

Details of the specific single-cell ionic current models we use 
are provided further below.

\section{NUMERICAL METHODS OF SOLUTION}
\subsection{Operator splitting}
The monodomain model \eqref{eq::monodomain} is characterised by a
large range of significant scales, e.g. cardiac
action potentials have extremely fast and narrow upstrokes
(depolarization) and very slow and bread recovery (repolarization)
phases. An effective numerical scheme based on an operator splitting
approach (Godunov and Strang splitting, \citep{Strang1968}, also known
as the fractional timestep method  \citep{Press92a}), was
proposed by \citet{ZhilinQu1999} and is adopted in our study  in the following form.
The nonlinear monodomain model \eqref{eq::monodomain} is split into a
set of nonlinear ordinary differential equations
\begin{subequations}
\begin{align}
    \frac{\partial V}{\partial t} &=
    -\frac{1}{C_m}\big(I_\text{ion}(V,\vec{y}) + I_\text{stim}\big), \nonumber\\
\frac{\partial \vec{y}}{\partial t} &= \vec{R}\big(V,\vec{y}\big),
\label{eq::monodomain:cell}
\end{align}
and  a linear diffusion partial differential equation
\begin{align}
    \frac{\partial V}{\partial t} &= \frac{1}{\chi C_m}\nabla \cdot (\vec{\sigma} \nabla V).
\label{eq::monodomain:diff}
\end{align}
\end{subequations}
To integrate the complete monodomain model \eqref{eq::monodomain}
in the interval $[t_n,t_n+\Delta t]$ we take the following three fractional steps of the splitting algorithm.
\begin{enumerate}
    \item Solve the nonlinear ODE system for $V_\theta^{n}$ at $t_n < t \le t_n + \theta \Delta t$ with known $V^n$
      \begin{align*}
        \frac{\partial V}{\partial t} =
        -\frac{1}{C_m}I_\text{ion}(V,\vec{y}), 
        \qquad 
        \frac{\partial \v{y}}{\partial t} = \vec{R}(V, \vec{y}), 
        \qquad 
        V(t_n) = V^n. 
      \end{align*}
    
    \item Solve the linear PDE for $V_\theta^{n+1}$ at $t_n < t \le t_n + \Delta t$
        \begin{equation*}
           \frac{\partial V}{\partial t} = \frac{1}{\chi C_m}\nabla \cdot (\vec{\sigma} \nabla V), \quad \quad V(t_n) = V_\theta^{n}.
        \end{equation*} 
        
    \item Solve the ODE system again for $V^{n+1}$  at  $t_n + \theta \Delta t < t \le t_n +  \Delta t$
      \begin{align*}
        \frac{\partial V}{\partial t} =    -\frac{1}{C_m}I_\text{ion}(V,\vec{y}), 
        \qquad 
        \frac{\partial \v{y}}{\partial t} = \vec{R}(V, \vec{y}), 
        \qquad 
        V(t_n+ \theta \Delta t) = V_\theta^{n+1}.
      \end{align*}
\end{enumerate}
Further details on the operator splitting method applied to the
monodomain problem can be found in \citep{Sundnes2006}.

\subsection{Reaction part}
In this form the normally stiff initial value problem
\eqref{eq::monodomain:cell} can be integrated separately using 
one of the many known methods for solution of initial value problems,
including adaptive time stepping. Depending on the specific ionic
models,  a forward Euler method may be used for temporal stepping for
less stiff cases, or a fourth-order Runge-Kutta method, for stiffer
problems, for instance. 

\subsection{Diffusion part}
The diffusion part of the  monodomain model \eqref{eq::monodomain} is solved using a
finite-element method as detailed below. 
For the spacial disretisation of the equation 
\eqref{eq::monodomain:diff} the numerical approximation $V^h(\vec{x},t)$ of the transmembrane
voltage potential $V(\vec{x},t)$ is assumed to take the form of 
a finite expansion in a set of continuous piecewise polynomial nodal
basis functions $\big\{\phi^h_i(\vec{x}), i=1..p\big\}$ with
time-dependent coefficients $\big\{V_i(t), i=1..p\big\}$ each
representing a nodal value at time $t$ 
\begin{align}
V(\vec{x},t) \approx V^h(\vec{x},t) 
= \sum_{i=1}^p \phi_i(\vec{x}) V_i(t),
\label{eq::femexpansion}
\end{align}
where $p = \dim\{\phi^h\}$, and $h$ denotes a parameter measuring the
size of the domain partition. 
Substituting expansion \eqref{eq::femexpansion} in equation
\eqref{eq::tissue_monodomain}, taking the Galerkin projection and 
using the boundary condition \eqref{eq::bc_monodomain}, the following
weak variational form of the monodomain equation is obtained
\begin{align}
\label{eq::variational_form_mon}
   \chi \, C_m \Big(\frac{\partial V^h}{\partial t}, \phi^h_i\Big)_\Omega +
  \Big(\vec{\sigma} \nabla V^h, \nabla \phi^h_i\Big)_\Omega =0 , \\
\quad i=1,\dots,p,  \nonumber
\end{align}
representing a weighted-residual condition for minimization of the
residual error, where the round brackets $(v,w)_\Omega
= \int_{\Omega} vw \,d\Omega$ denote the inner product with the basis functions.
The Galerkin approximaiton \eqref{eq::variational_form_mon} represents
a set of $p$ ordinary differential equations in time for the
$p$ coefficient functions $V_i(t)$ in the expansion \eqref{eq::femexpansion}.
For brevity in the following we will drop the superscript $h$.

For the temporal disretisation of the Galerkin projection equations
\eqref{eq::variational_form_mon} the time derivative approximated a
first-order accurate forward finite difference formula and 
the following implicit numerical scheme is used
\begin{equation}
\chi \, C_m \, \vec{M}\frac{\vec{V}^{n+1}-\vec{V}^n}{\Delta t} + \vec{K}\,\vec{V}^{n+1} = 0,
\label{eq::algberiac_fe_mono}
\end{equation}
where $\vec{V}^n = [V_1^n,V_2^n,\dots, V_p^n]^T$ now denotes
$p$-dimensional vector of voltage values at time level $t_n=n\Delta
t$ with time step $\Delta t$, and where 
$$
[M_{ij}] = (\phi_i, \phi_j)_\Omega,
$$
denotes the mass matrix and
$$
[K_{ij}] = (\nabla \phi_i, \vec{\sigma}\nabla \phi_j)_\Omega,
$$
denotes the stiffness matrix. Finally, the vector of unknowns voltage
values at time level $t_{n+1}$ is determined by solving the linear
system of equations
\begin{equation}
(\vec{M} + \frac{\Delta t}{\chi \, C_m}\, \vec{K}) \, \vec{V}^{n+1} =  \vec{M}\,\vec{V}^n.
\label{eq::monodomain_FE}
\end{equation}

\subsection{Practical implementation}
Equation \eqref{eq::monodomain_FE} is solved using the
\texttt{libMesh} open source parallel C++ finite element
library\footnote{\url{libmesh.github.io}} \citep{Kirk2006}, and the solution of linear
systems and the time stepping relies on the solvers provided by the
\texttt{PETSc} library\footnote{\url{www.mcs.anl.gov/petsc}}.
Simulations are run both on our local Linux workstations with 2
Intel(R) Xeon (R) CPU E5-2699 2.30 GHz (up to 72 threads) and 128 GB
of memory at the School of Mathematics and Statistics, University of
Glasgow as well as on the RCUK flagship High-Performance parallel
computer
ARCHER\footnote{\url{www.archer.ac.uk}}. Visit\footnote{\url{https://visit.llnl.gov}}
is used for post-processing the two- and three-dimensional simulations. 

\begin{figure}[t]
	\centering
	\includegraphics[width=0.75\linewidth]{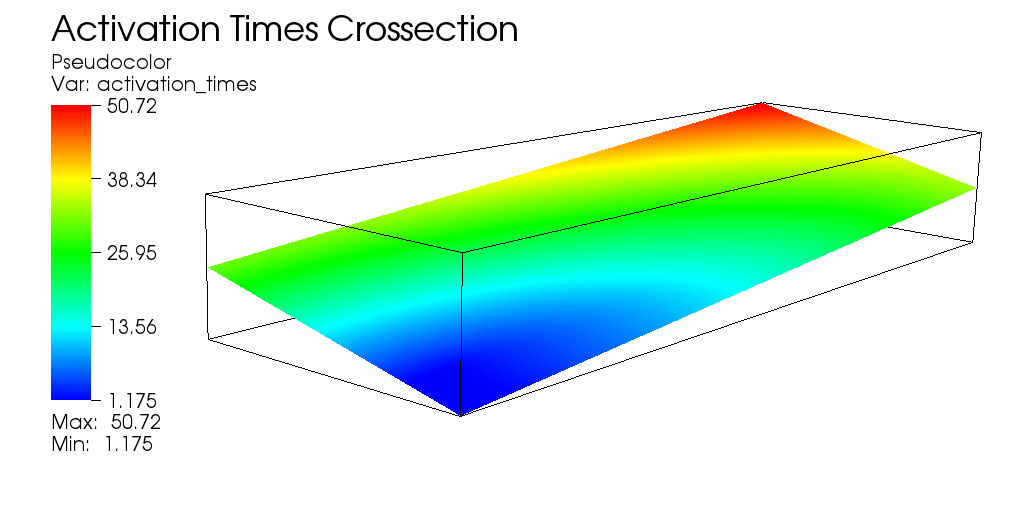}
\caption{Computational domain
A contour plot of activation times for the benchmark problem defined
in section \ref{sect::benchmark}. The computational domain is clearly
visible. The stimulus site is in the lower central vertex while the
most distant point is the upper central vertex.
}
\label{fig::bench:geometry}
\end{figure}

\begin{table}[t]
	\centering
	\begin{tabular}{c l |c c c c|} 
		\cline{3-6}
		~ & ~ & \multicolumn{4}{c|}{$\Delta x$}\\
		\cline{3-6}
		~ & ~ & 0.5mm & 0.333mm & 0.2mm & 0.1mm \\ [0.5ex] 
		\hline
		\multicolumn{1}{ |c | }{\multirow{4}{*}{$\Delta t$}}&0.05ms & 81.75 & 60.95 & 52.15 & 47.20 \\ 
		\multicolumn{1}{ |c | }{}	&0.025ms & 80.70 & 59.85 & 49.94 & 45.40\\
		\multicolumn{1}{ |c | }{}	&0.01ms  & 80.06 & 59.20 & 49.94 & 44.26\\
		\multicolumn{1}{ |c | }{}	&0.005ms & 79.82 & 58.96 & 49.65 & 43.85\\
		\hline
	\end{tabular}
	\caption{Values for the activation time [ms] at different spatial and
          temporal discretisation steps measured in our numerical code for the benchmark
          problem described in section \ref{sect::benchmark}.} 
	\label{table:TPBench}
\end{table}

\begin{figure}[t]
\centering
 \includegraphics[width=0.99\linewidth]{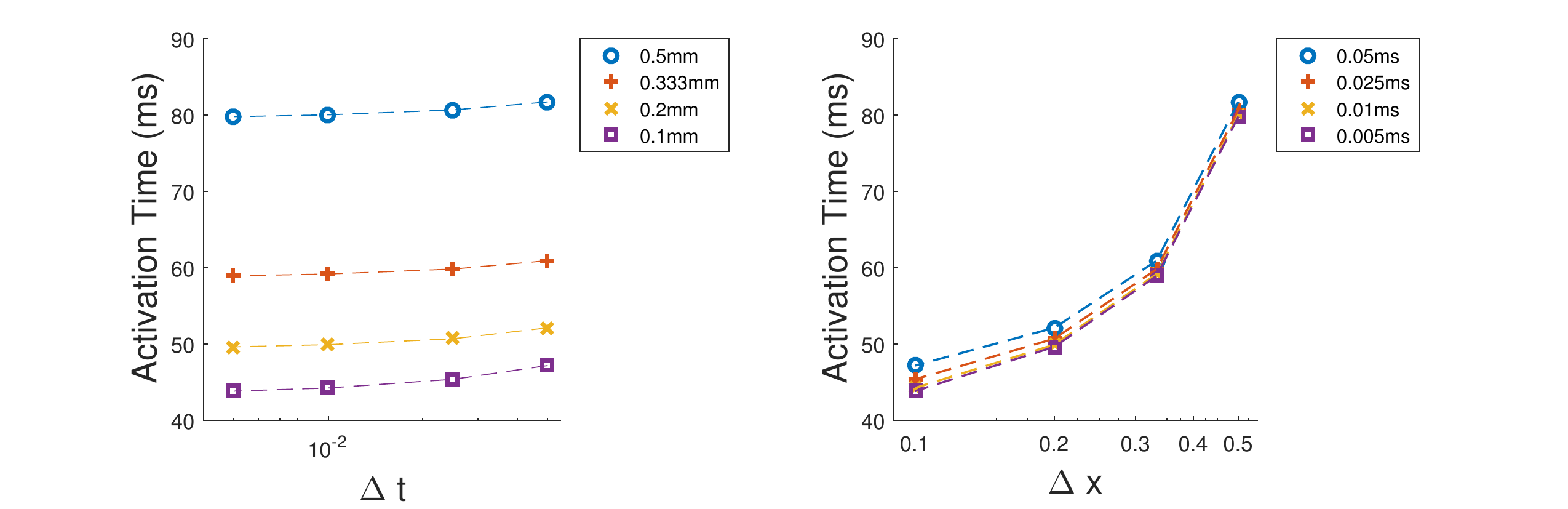}
\caption{Convergence of the value of the activation time as time and
  space steps are decreased. Values are measured in our numerical code
for the benchmark problem described in section \ref{sect::benchmark}.}
\label{fig:tpbenchlog}
\end{figure}

\section{BENCHMARKING AND VALIDATION}
\label{sect::benchmark}

\subsection{Benchmark description}
Our mathematical model and its numerical implementation was validated
by comparison with a standard cardiac tissue electrophysiology simulation
benchmark case developed by the research community \citep{Niederer2011}.
The benchmark involved 11 independently developed numerical codes
providing numerical simulations of a well-defined problem with unique
solution for a number of different resolutions. The benchmark seeks to
compare solutions of the monodomain equations \eqref{eq::monodomain}
on a cuboid domain of dimensions $20 \times 7 \times 3$ mm
using the
\citet{tenTusscher2006} model of human epicardial myocytes as a model
of the transmembrane ion current density $I_\text{ion}$. The initial
stimulus current $I_\text{stim}$ has a current density amplitude of
$50000 \mu\text{A cm}^{-3}$ and is applied to a cube with size $1.5
\times 1.5 \times 1.5$ mm positioned at the corner of the full cuboid
domain and a stimulus duration of 2 ms. The value of the cell surface
to volume ratio $\chi$ is 140 mm$^{-1}$, and it is assumed that the
cardiac fibres are aligned with the long, 20 mm, axis of the cuboid
domain so the conductivity tensor $\vec{\sigma}$ is diagonal with
values $[0.1334, 0.0176, 0.0176]$  S m$^{-1}$ along its main
diagonal. The so called ``activation time'' defined as the time it
takes for a cardiac action potential to travel from the stimulation
site to the most distant point in the computational domain (i.e. the
point opposite the stimulation site) is requested as a diagnostic
output quantity from the numerical simulation.  Figure
\ref{fig::bench:geometry}  shows the geometry of the benchmark case.

\subsection{Validation and benchmarking}
We have verified that our numerical code is in excellent agreement
with the community benchmark results. Since the computational domain
of the benchmark problem is rectangular domain we have used a regular
square finite-element mesh with space step $\Delta x$. For
time stepping we have used the simple forward Euler method with
time step $\Delta t$. Table \ref{table:TPBench} shows the values of
the activation time we have obtained at various spatial and temporal
resolutions. At the highest resolution of $\Delta x=0.1$ mm and
$\Delta t = 10^{-4}$ ms  the activation time obtained using our code
is $43.85$ ms which is within 
2\% error bar from the 42.82 ms high-accuracy value agreed upon in the
benchmark paper \citep{Niederer2011}.  Figure \ref{fig:tpbenchlog}
shows a convergence test we have performed with decreasing space step
and time step and it is clear that our solution is converging to
values closer to the community benchmark value just quoted. We remark
that for our code the increase of the spatial resolution leads to more
significant increases of accuracy than the increase in temporal
resolution.
\begin{table}[t]
\centering
 \begin{tabular}{l c c} 
\hline 
 Parameters & M-cells (Healthy) & M-cells (MI)    \\  \hline 
 $K_0$, External potassium concentration (mM) &4.5865 &4.3087  \\  
 $Ca_0$, External calcium concentration (mM)&2.0467&2.5678 \\ 
 $Na_0$, External sodium concentration (mM) &146.30 &165.71  \\ 
 $gna$, Peak INa conductance & 14 & 12  \\ 
 $gca$, Strength of Ca current flux (mmol/(cm C)) &259.86 & 193.23 \\  
 $pca$, Constant in ICal (cm/s) &0.0002 & 0.0008 \\ 
 $r1$, Opening rate in ICal & 0.4804 &0.3546  \\ 
 $r2$, Closing rate in ICal &  2.2825 &2.8694    \\ 
 $gkix$, Peak IK1 conductance (mS/$\mu$F) &  0.4400 & 0.2604     \\ 
 $gtof$, Peak Ito conductance (mS/$\mu$F) &0.0221 & 0.0421 \\
 \hline
 (Global) root mean square error &  0.0133 & 0.1070 \\      [1ex] 
\hline
\end{tabular}
\caption{Parameter values of the model of \citet{Mahajan2008} re-fitted to
  the experimental data on M-cells at 3 Hz pacing rate form \citep{McIntosh2000}.}
\label{table3}
\end{table}
\begin{figure}[t]
\centering
\includegraphics[width=0.5\textwidth]{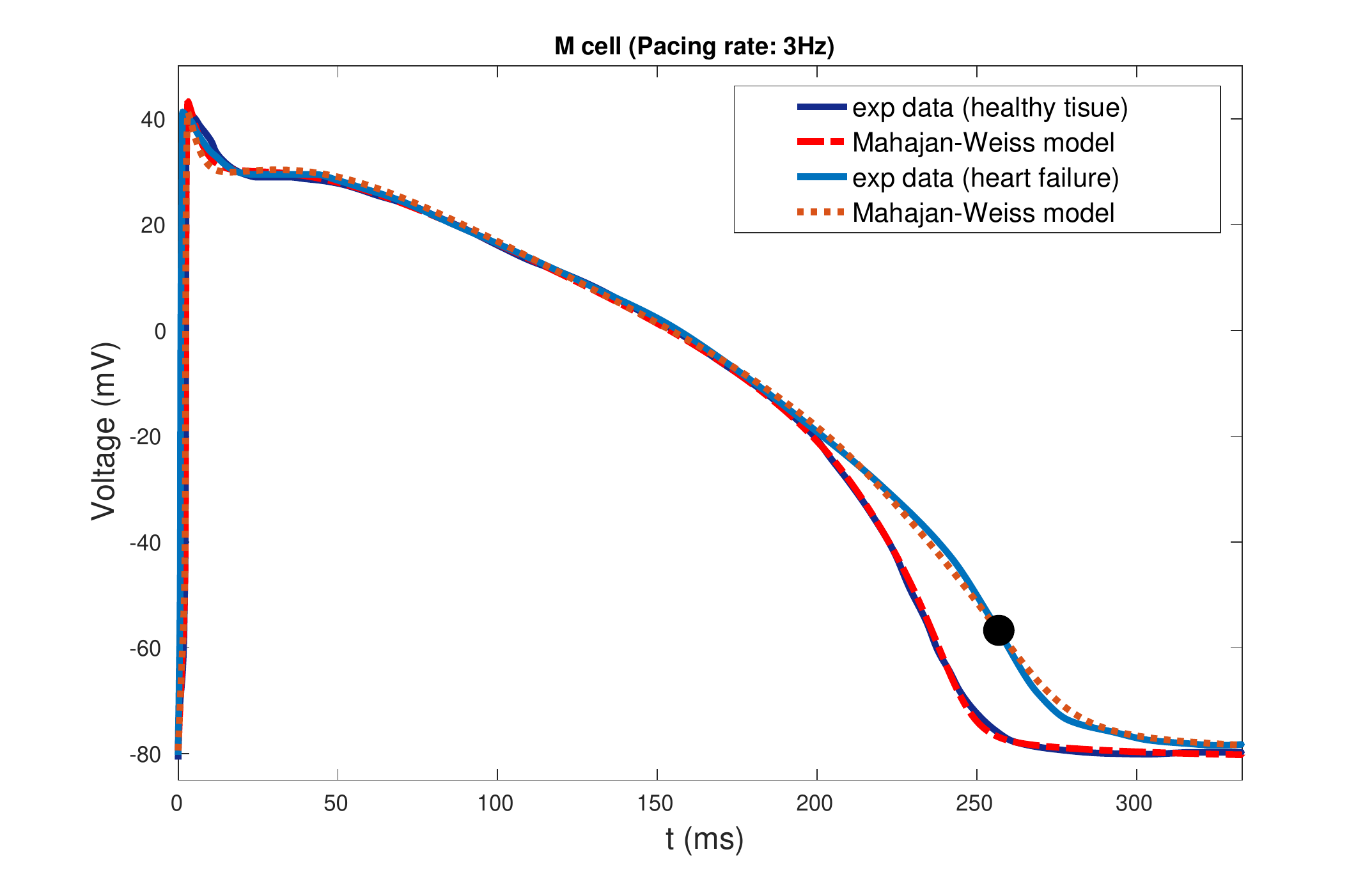}
\caption{Action potential computed using the model of \citet{Mahajan2008} with
  parameter values given in table \ref{table3} in comparison with
  experimental data from \citep{McIntosh2000}.} \label{AP_Mcell}
\end{figure}

\section{PARAMETER RE-FITTING OF A RABBIT VENTRICULAR SINGLE CELL
  IONIC MODEL}
\label{sect::refitting}
     
In order to achieve an accurate comparison with experimental
measurements in rabbit ventricular tissue samples
e.g. \citep{Allan2016,Myles2010,Myles2009phd} an appropriate single 
cell ionic action potential model must be selected and refitted.
To this end we have selected to use \citet{Mahajan2008} detailed
action potential model, one of the modern rabbit ventricular AP models
designed to accurately reproduce the dynamics of the cardiac action
potential and intracellular calcium (Cai) cycling at rapid heart rates
as relevant to ventricular tachycardia and fibrillation. 
Cardiac electrophysiology models are based on
experimental data from a variety of sources, including measurements in
different species and under different experimental conditions
\citep{Cooper2016}. Refitting of model parameters is therefore
necessary whenever new or more appropriate data sets are available. 
In our case, the model of \citet{Mahajan2008} was refitted to match the
single cell experimental data of  \citet{McIntosh2000} since these
were measured by the same research group using identical experimental protocols.

In \citep{McIntosh2000} action potential and intracellular Ca2+
transient characteristics $x_i^\text{target}$ were measured in single
cardiac myocytes from mid-myocardial regions of the left ventricle of
rabbits with and without heart failure.   These were fitted to the
outputs of the model of \citet{Mahajan2008}, $x_i^\text{sim}$, by
minimising the  error function 
\begin{equation} 
\label{eq:err}
\text{Err}_{\text{AP}} = \frac{1}{M} \sum_{i=1}^M  (x_i^\text{sim} - x_i^\text{target})^2
\end{equation}
with respect to selected parameter values, aka ``parameter estimation''.
For parameter estimation we used  a standard Matlab routine for
unconstrained multivariable minimisation based on the  bounded
Nelder-Mead simplex-like method \citep{Lagarias}. The results are
shown in Table \ref{table3} and figure \ref{AP_Mcell} below.
\begin{table}[t]
\centering
\begin{tabular}{c l |c c c c |} 
		\cline{3-6}
		~ & ~ & \multicolumn{4}{c|}{$\Delta x$}\\
		\cline{3-6}
		~&~ & 0.5mm & 0.333mm & 0.2mm & 0.1mm \\ [0.5ex] 
		\hline
		\multicolumn{1}{ |c | }{\multirow{5}{*}{$\Delta t$}}&0.05ms & X & X & X & X \\ 
		\multicolumn{1}{ |c | }{}&	0.01ms   & X & X    & X     & 54.19 \\
		\multicolumn{1}{ |c | }{}&	0.005ms  & X & X    & 63.94 & 53.90 \\
		\multicolumn{1}{ |c | }{}&	0.0025ms & X &82.30 & 63.81 & 53.76 \\
		\multicolumn{1}{ |c | }{}&	0.0001ms & X &82.25 & 63.75 & 53.68 \\
		\hline
\end{tabular}
\caption{Values for the activation time [ms] at different spatial and temporal
discretisation steps measured in our numerical code for the benchmark
geometry described in section 4 but for the model of \citet{Mahajan2008}
refitted to healthy values rather than \citet{tenTusscher2006} model.}
\label{table:WeissBench}
\end{table}
\begin{figure}[t]
\centering
\includegraphics[width=0.99\linewidth]{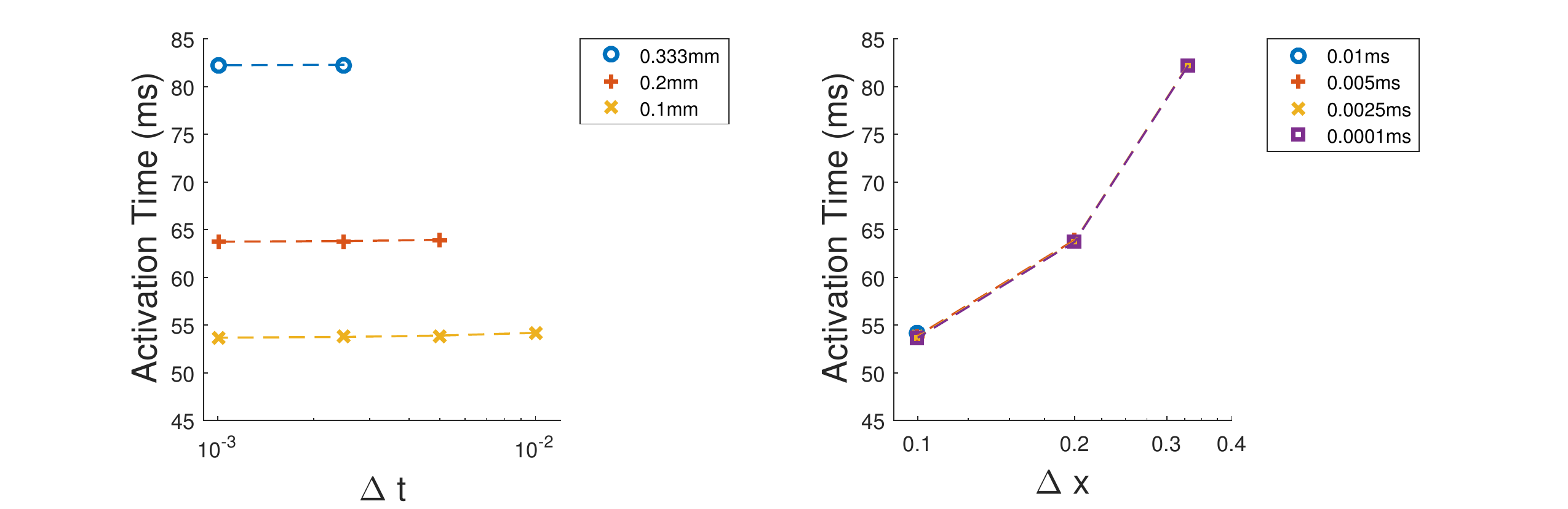}
\caption{
Convergence of the value of the activation time as time and
  space steps are decreased. Values are measured in our numerical code
for the benchmark problem described in section \ref{sect::benchmark}
with the model of \citet{Mahajan2008} refitted to healthy values.}
\label{fig:weissbenchlog}
\end{figure}

The benchmark convergence test of section \ref{sect::benchmark} was
repeated using the model of \citet{Mahajan2008} newly re-fitted to healthy
values in order to establish suitable resolution. Based on the results
of Table \ref{table:WeissBench} and figure \ref{fig:weissbenchlog} we
determine that $\Delta x= 0.1$ and $\Delta t = 5\times10^{-3}$ ms provides a good
trade-off between resolution and model accuracy and we use this values
for the simulations detailed in the next section.

\section{MODELLING OF PROPAGATION IN SCARRED TRANSMURAL VENTRICULAR SLABS}

\begin{figure}[t]
\centering
 \begin{tabular}{ll}
a) & b) \\
\includegraphics[width=0.22\linewidth]{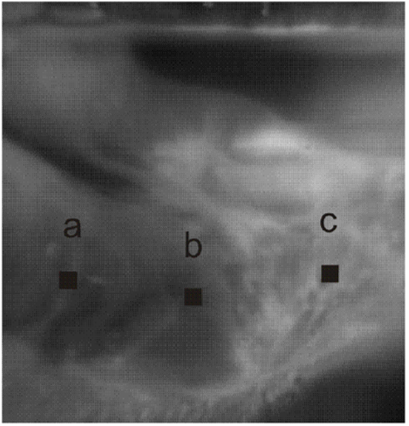}    &
\includegraphics[width=0.42\linewidth]{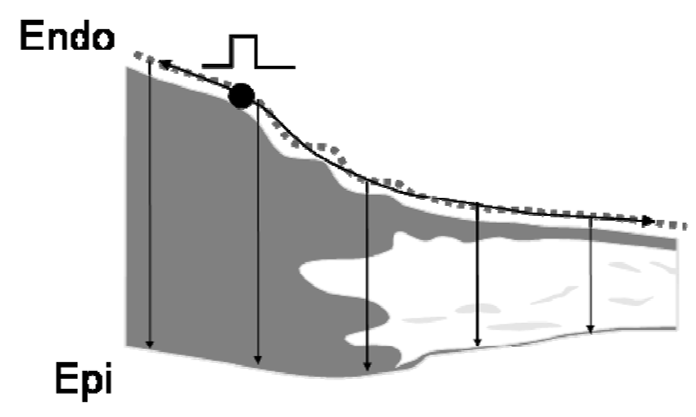}\\
c) & d) \\
\includegraphics[width=0.42\linewidth]{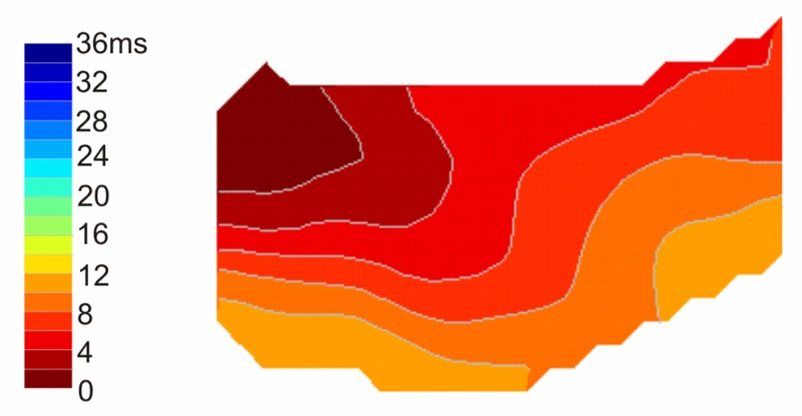}   &
   \includegraphics[width=0.42\linewidth]{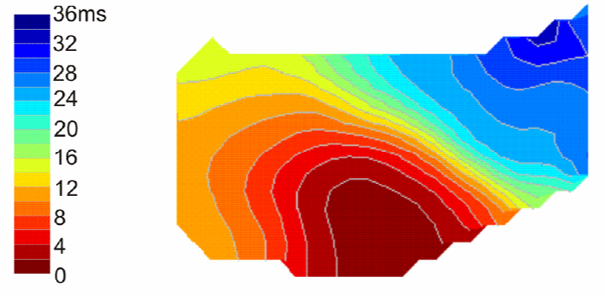}   \\
 \end{tabular}
\caption{Transmural conduction into an  infarct zone. Plots taken from figure 5.8 of \citep{Myles2009phd}.}
\label{fig::expdata}
\end{figure}
\begin{figure}[t]
  \centering
\begin{tabular}{ll}
(a) & (b) \\
  \includegraphics[width=0.4\linewidth]{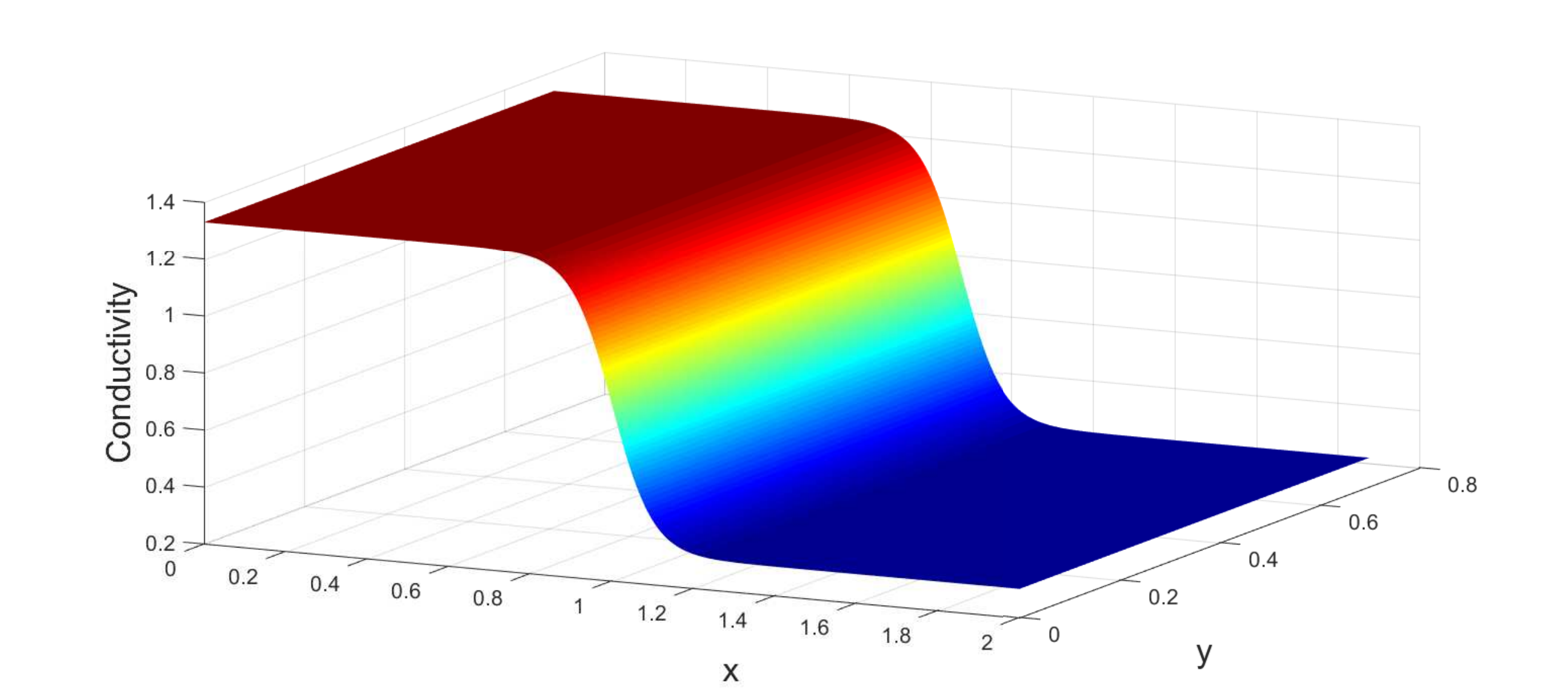}&
  \includegraphics[width=0.49\linewidth]{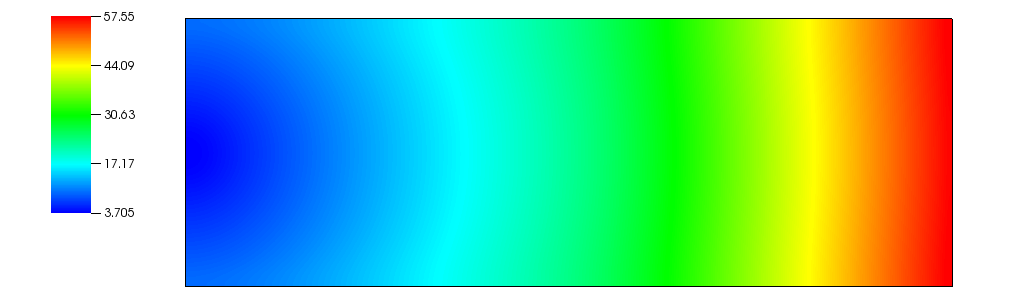}\\
\end{tabular}
  \caption{(a) Isotropic conductivity as a function of $x$ as given by
    equation \eqref{eq:cond:sigm}. (b) Simulated activation times [ms] for the conductivity
    profile in part (a).} 
  \label{fig:sig}
\centering
  \begin{tabular}{l l l}
    \multirow{2}{*}{\includegraphics[width=0.12\linewidth]{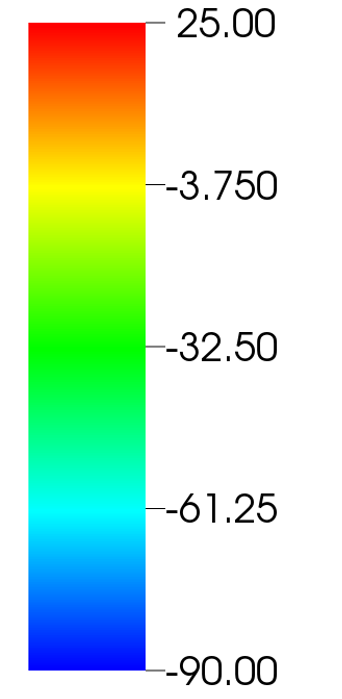}}    &
    \includegraphics[width=0.42\linewidth]{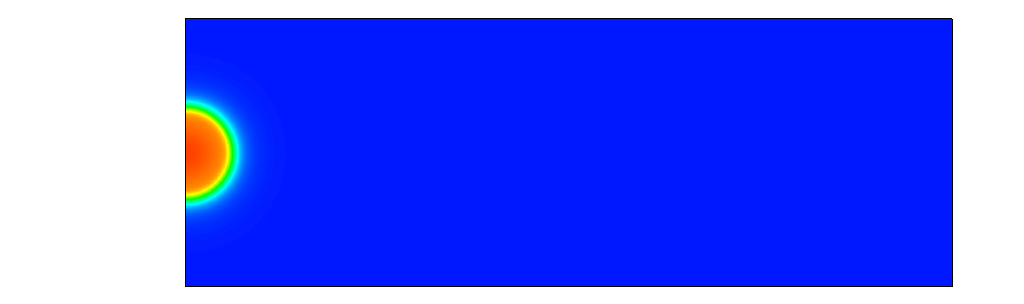}    &
    \includegraphics[width=0.42\linewidth]{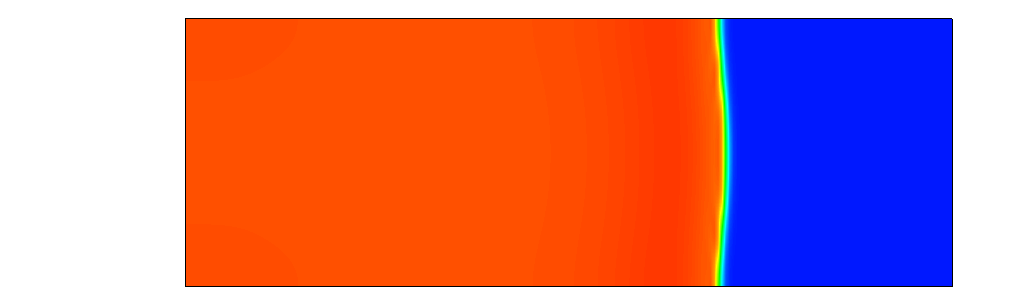}    \\
    ~&
    \includegraphics[width=0.42\linewidth]{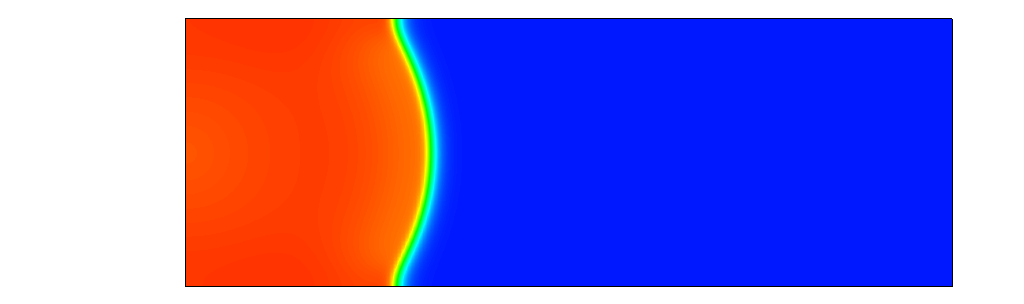}    &
    \includegraphics[width=0.42\linewidth]{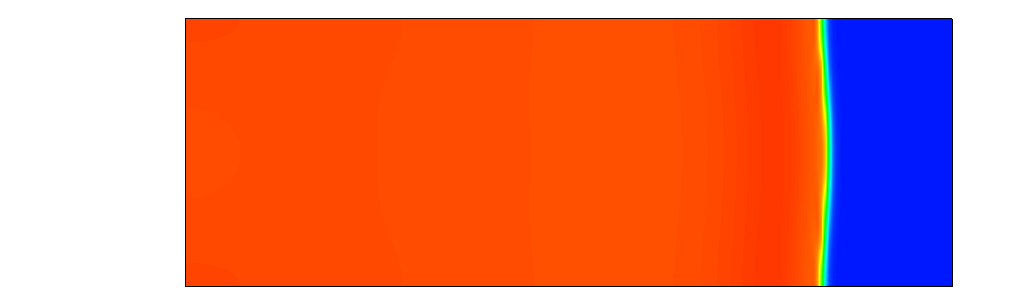}    \\
    ~&
    \includegraphics[width=0.42\linewidth]{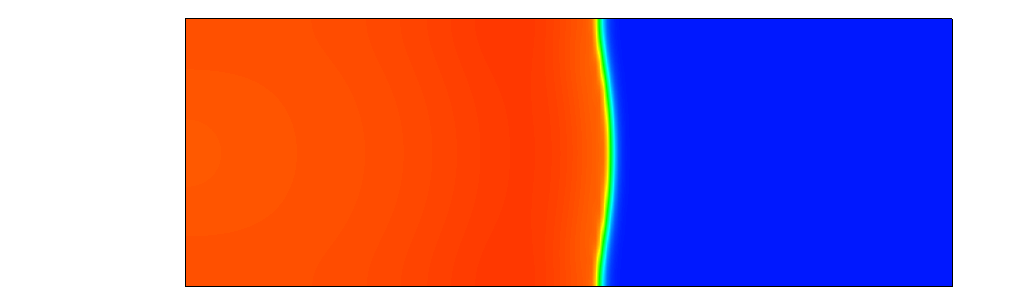}
    &
    \includegraphics[width=0.42\linewidth]{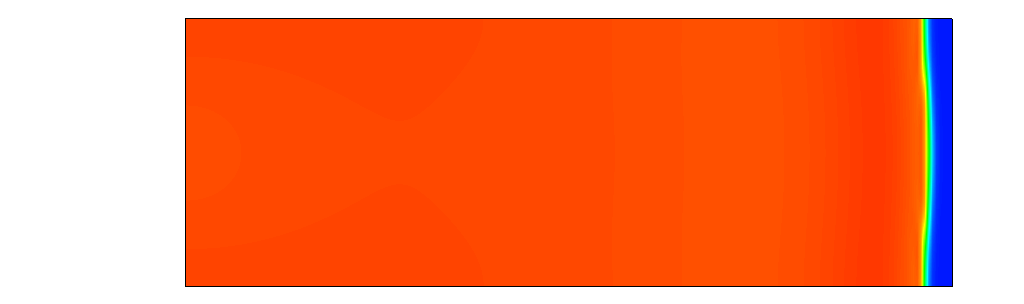}
  \end{tabular}
\caption{Propagation of an action potential in the case of sigmoidal
  conductivity \eqref{eq:cond:sigm}. Density maps of the transmembrane
  voltage potential are plotted at equidistant times $t_i= 5 + i \Delta
  t$,   $i=1,\dots, 6$ and $\Delta t= 10$ ms.}
\label{fig:voltages:sigm}
\end{figure}

\begin{figure}[t]
  \centering
  \begin{tabular}{ll}
  (a) & (b) \\
    \includegraphics[width=0.4\linewidth]{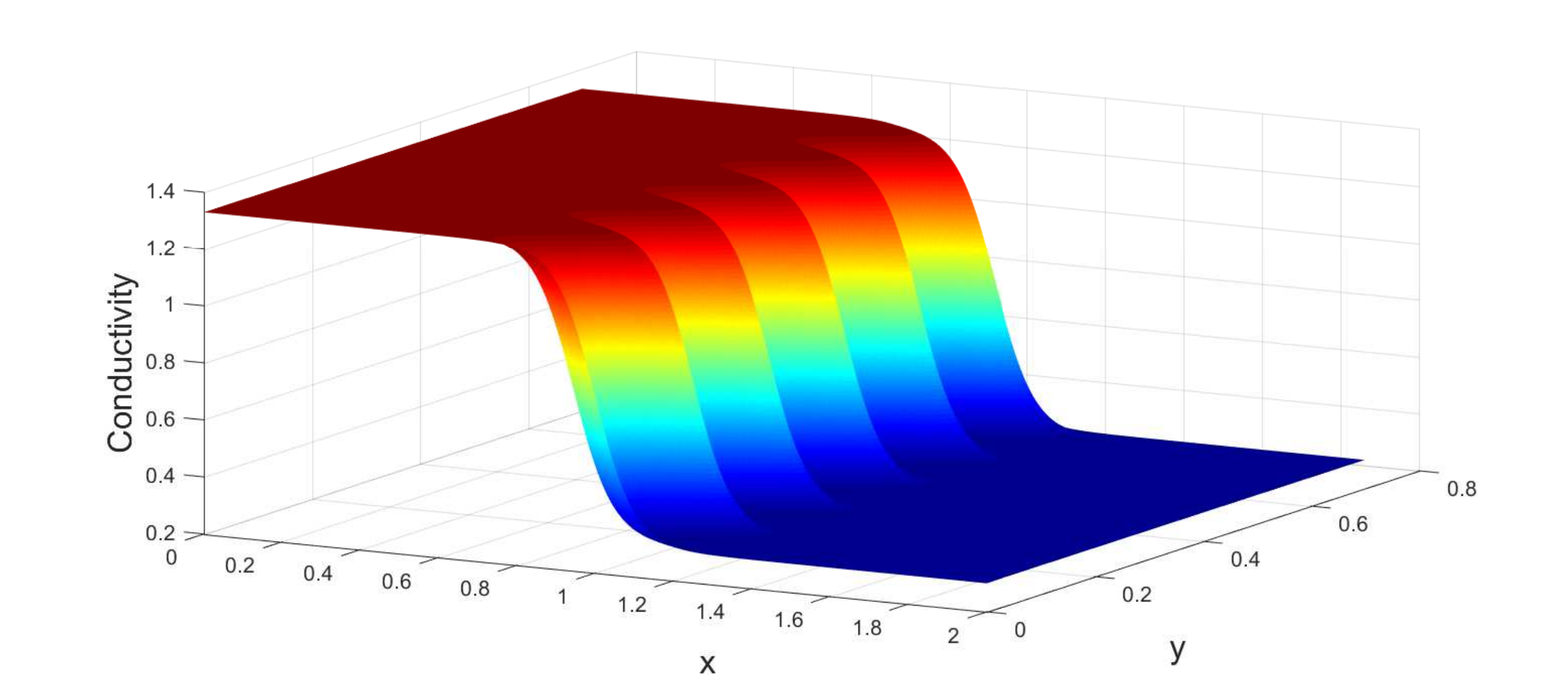} &
    \includegraphics[width=0.49\linewidth]{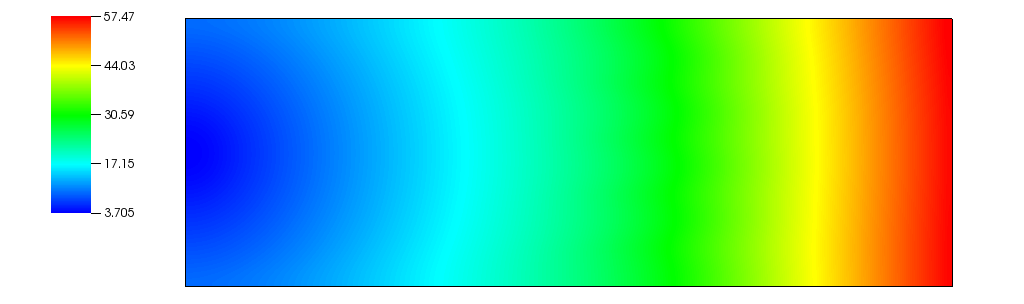}\\
  \end{tabular}
  \caption{(a) Isotropic conductivity with a fingering effect as a
    function of $x$ and $y$ as given by equation
    \eqref{eq:cond:fing}. (b) Simulated activation times [ms] for the
    conductivity     profile in part (a).} 
  \label{fig:sigsine}
\centering
\begin{tabular}{l l l}
  \multirow{2}{*}{\includegraphics[width=0.12\linewidth]{VLegend}}
  &
  \includegraphics[width=0.42\linewidth]{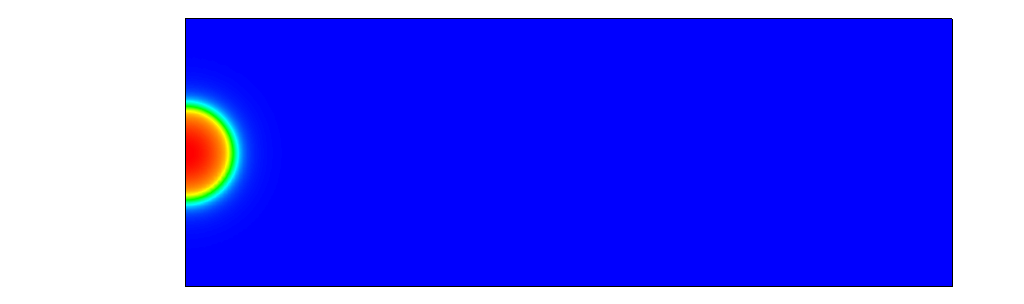} &
  \includegraphics[width=0.42\linewidth]{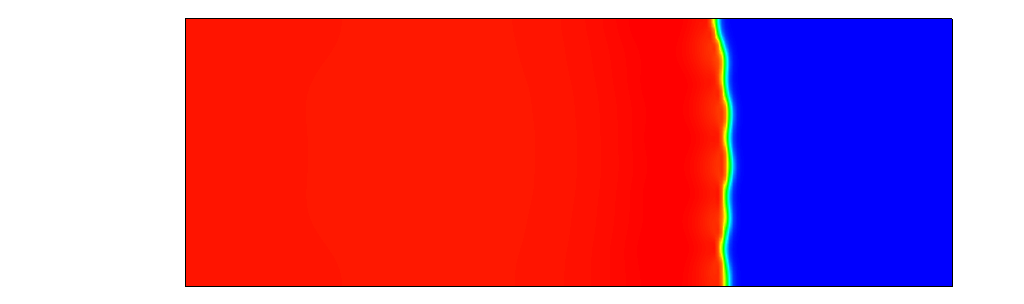} \\
  ~&
  \includegraphics[width=0.42\linewidth]{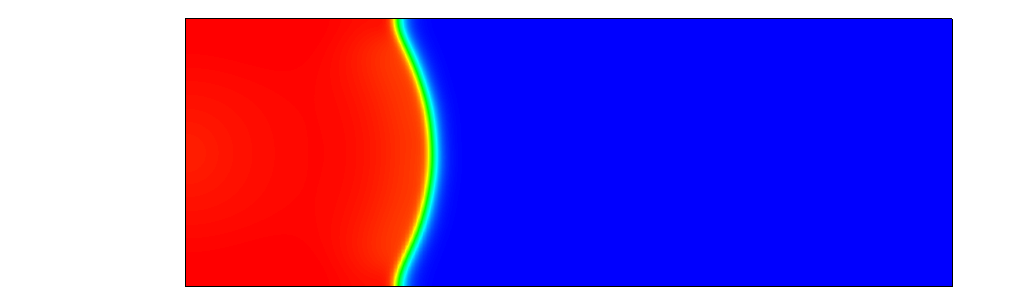} &
  \includegraphics[width=0.42\linewidth]{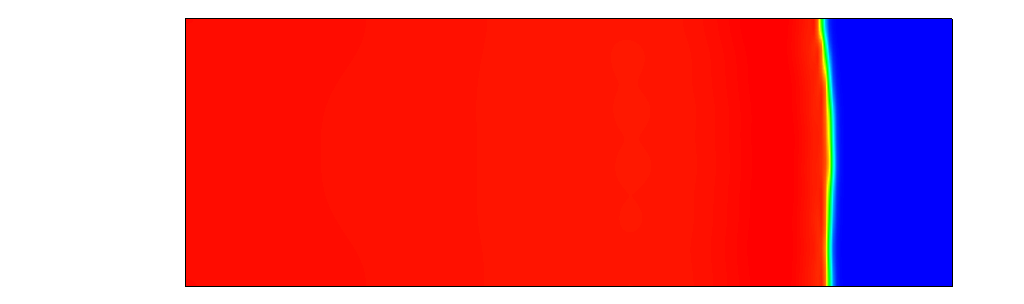} \\
  ~&
  \includegraphics[width=0.42\linewidth]{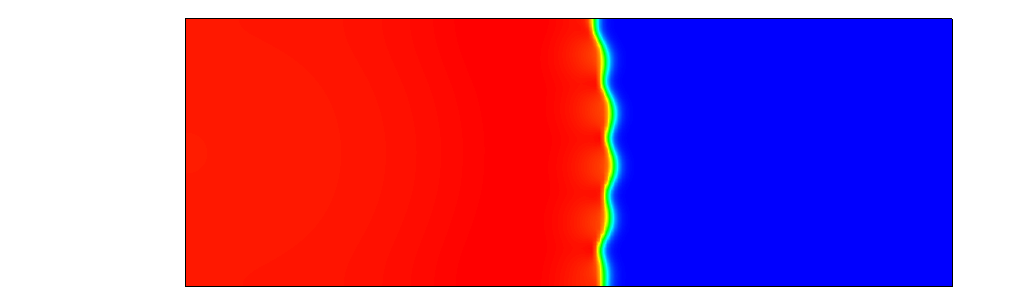} &
  \includegraphics[width=0.42\linewidth]{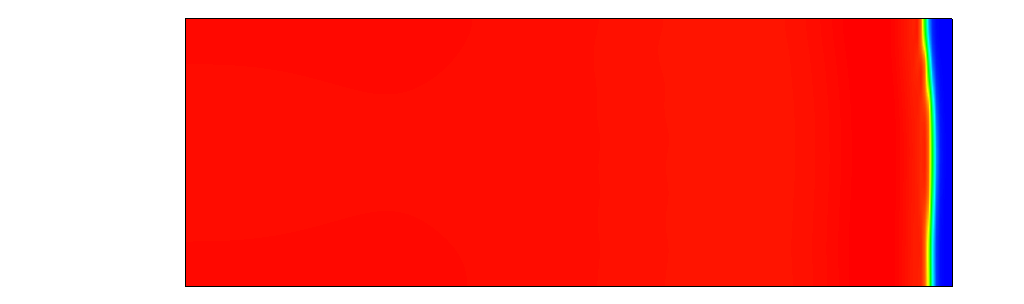}
\end{tabular}
\caption{Propagation of an action potential in the case of finger-like 
  conductivity \eqref{eq:cond:fing}. Density maps of the transmembrane
  voltage potential are plotted at equidistant times $t_i= 5 + i \Delta
  t$,   $i=1,\dots, 6$ and $\Delta t= 10$ ms.}
\label{fig:voltages:fing}
\end{figure}

\subsection{Physiology of the infarcted zone}

\subsection{Description of experiments}
An example of the experimental measurements of transmural conduction
into an infarct zone available from our collaborators
\citep{Myles2009phd} is shown in Figure \ref{fig::expdata}.
The upper left panel shows a plain image of the transmural surface of
a wedge preparation from a ligated heart, with the endocardium 
uppermost and the epicardium at the bottom of the picture. The black
squares indicate the position from which example APs are available for
comparison: a) remote zone b) border zone and c) infarct zone. 
In the upper right panel is a schematic diagram of the preparation,
indicating the position and the shape of the infarct border zone. 
The lower two panels show isochronal maps of activation time during  
endocardial and epicardial stimulation at left and right panels, respectively.
The experiment has a number of notable features, including,
\begin{description}
\item[a)] The infarct zone has lower density of electrically excitable
  cells. 
\item[b)] The infarct border zone has significant undulations that
  protrude the healthy zone. 
\item[c)] The infarct zone has a reduced volume compared to the
  healthy zone resulting in a wedge-like trapezoidal shape of the
  transmural slab rather than a rectangular shape.
\end{description}
We will take the approach of modelling these features separately, in
order to investigate their effects one at a time before we attempt to
address the phenomena in full complexity and detail. 
To this end we perform direct numerical simulations of the monodomain
problem \eqref{eq::monodomain} as specified in section
\ref{sect::benchmark} except that the ionic model $I_\text{ion}$ is
replaced by the model of \citet{Mahajan2008} refitted to healthy values
as described in section \ref{sect::refitting} and conductivity values
as specified further below.

\subsection{Modelling}
The simplest way to model the infarct zone and feature (a) is to 
assume that the lower density of the myocites in the infarct can be
described by a reduced effective value of the conductivity in the
infarct zone. To further
focus on the effect of a well-defined border zone we will also assume
that conductivity is isotropic so all components or the conductivity
tensor are equal to the same scalar value $\sigma$. We take this value
to depend sigmoidally on the intra-longitudinal coordinate $x$,
\begin{equation}
  \sigma(x) = \sigma_a +(\sigma_b - \sigma_b)\frac{\exp(\alpha
    (x-x_0))}{1+\exp(\alpha(x-x_0))}, 
\qquad 
x_0=1,
\label{eq:cond:sigm}
\end{equation}
where $\sigma_a=1.3342$ and $\sigma_b=0.3$ are the values of the conductivity
deep into the healthy and the infarct zone, respectively, $x_0$ is the
location of the border zone, $\alpha=10$ is the steepness of the
sigmoidal function. This conductivity profile is shown in
\ref{fig:sig}(a). The corresponding activation times are shown in
Figure \ref{fig:sig}(b) while snapshots of the transmembrane voltage
potential at a set of equidistant moments are shown in Figure
\ref{fig:voltages:sigm}.  
The simulations show that the travelling front propagates faster when
the conductivity is large and slows down when the conductivity is
small. This effect is not observed in the experimental measurements
as seen in both Figures \ref{fig::expdata}(a,b). 

Figure \ref{fig::expdata}(b) shows in fact that the propagating wave
slows down within the infarct border zone but subsequently speeds up
when in the infarct zone and travels to as a speed similar to the 
speed in the healthy zone. To investigate if this is an effect of the
finger-like undulations in the infarct border zone we consider a
conductivity profile given by the expression
\begin{equation}
  \sigma(x,y) = \sigma_a +(\sigma_b - \sigma_b)\frac{\exp(\alpha
    (x-x_0(y)))}{1+\exp(\alpha(x-x_0(y)))},  
\qquad         
x_0(y) =  1+0.1\sin(44.86y), 
\label{eq:cond:fing}
\end{equation}
where the border location $x_0$ is now modulated as a function of the
intra-transversal $y$-direction. The modulating sine function mimics a 
fingering effect as shown in Figure \ref{fig:sigsine}(a). The
corresponding activation times are shown in Figure
\ref{fig:sigsine}(b) while snapshots of the transmembrane voltage
potential at a set of equidistant moments are shown in Figure  \ref{fig:voltages:fing}.
The simulation results in this case are rather similar to the case of
unmodulated infarct boundary apart from a weak modulation of the
action potential front when it passes through the border. No speed-up
is observed within the infarct zone.

\section{CONCLUSION}

We have constructed a mathematical model and implemented a numerical
code for the solution of the monodomain problem \ref{eq::monodomain} for the
description of propagation of electrical excitation in cardiac tissue.
We have validated the code against a community developed benchmark.
We have selected an appropriate single cell ionic current model and
we have re-fitted its parameters to experimental data that conforms
to the protocols and procedures used in the lab of our collaborators.
With this we have performed several direct numerical simulations where
an infarct zone is modelled simply as a region with reduced values of
the conductivity. This alone has not been sufficient to provide a
good qualitative comparison with observations even if the undulation
of the infarct border zone is taken into account. Our work can be
extended and refined in a number of ways. Firstly, the conductivity
values in the healthy and the infarct zones can be better estimated by
further parameter fitting, this time applied to the spacially extended
problem. Secondly, the parameters of the conductivity profiles should
be systematically investigated. The wedge-like shape of the
experimental tissue sample should be taken into account.  The
model of \citet{Mahajan2008} re-fitted to heart-failure values should be
used within the infarct zone. These and further features will be
considered systematically in our future work.

\section*{ACKNOWLEDGEMENTS}

This work was supported by the EPSRC grant EP/N014642/1 ``SofTMech centre for
Multiscale Soft Tissue Mechanics with applications to heart and cancer''.


\end{document}